\documentclass[letterpaper, 10 pt, conference]{ieeeconf}  

\usepackage{graphicx}
\usepackage{amsmath,amssymb,amsfonts}
\usepackage{multirow}
\usepackage{algorithm}
\usepackage{algorithmic}
\usepackage{psfrag}
\usepackage{subfigure}
\usepackage{color}
\usepackage{soul}
\usepackage{xcolor}
\usepackage[hyphens]{url}
\usepackage[bookmarks=false]{hyperref}
\usepackage[hyphenbreaks]{breakurl}
\usepackage{lipsum}
\usepackage{mathrsfs}
\usepackage{pbox}
\usepackage{flushend}
\usepackage{mathtools}
\usepackage{tabularx}
\usepackage{caption}
\usepackage{url}
\usepackage{wrapfig}

\makeatletter
\def\hlinewd#1{%
  \noalign{\ifnum0=`}\fi\hrule \@height #1 \futurelet
   \reserved@a\@xhline}
\makeatother

\graphicspath{{Figures/}}

\IEEEoverridecommandlockouts                              

\title{\LARGE \bf SmartPathfinder: Pushing the Limits of Heuristic Solutions for Vehicle Routing Problem with Drones Using Reinforcement Learning
}

\author{Navid Mohammad Imran and Myounggyu Won
\thanks{Navid Mohammad Imran and Myounggyu Won are with the Department of Computer Science, University of Memphis, Memphis, TN, United States {\tt\small \{nimran, mwon\}@memphis.edu}}%
}

\begin{document}

\maketitle
\thispagestyle{empty}
\pagestyle{empty}

\begin{abstract}
The Vehicle Routing Problem with Drones (VRPD) seeks to optimize the routing paths for both trucks and drones, where the trucks are responsible for delivering parcels to customer locations, and the drones are dispatched from these trucks for parcel delivery, subsequently being retrieved by the trucks. Given the NP-Hard complexity of VRPD, numerous heuristic approaches have been introduced. However, improving solution quality and reducing computation time remain significant challenges. In this paper, we conduct a comprehensive examination of heuristic methods designed for solving VRPD, distilling and standardizing them into core elements. We then develop a novel reinforcement learning (RL) framework that is seamlessly integrated with the heuristic solution components, establishing a set of universal principles for incorporating the RL framework with heuristic strategies in an aim to improve both the solution quality and computation speed. This integration has been applied to a state-of-the-art heuristic solution for VRPD, showcasing the substantial benefits of incorporating the RL framework. Our evaluation results demonstrated that the heuristic solution incorporated with our RL framework not only elevated the quality of solutions but also achieved rapid computation speeds, especially when dealing with extensive customer locations.
\end{abstract}


\section{Introduction}
\label{sec:introduction}

The adoption of drone delivery systems has attracted considerable attention due to its numerous benefits. Drones transcend traditional delivery barriers, offering freedom from the constraints of road networks and avoiding traffic congestion, thus enhancing efficiency and reliability in parcel delivery~\cite{agatz2018optimization, lee2022congestion}. Furthermore, drone delivery has potential to significantly reduce CO2 emissions, aligning with global sustainability goals~\cite{goodchild2018delivery, chiang2019impact}. Additionally, the implementation of drones in delivery systems can lead to substantial cost reductions, primarily through the elimination of the need for truck drivers, which contributes to financial savings for logistics companies~\cite{dorling2016vehicle, pugliese2020using}. 

These compelling advantages have prompted a wave of investment from logistics companies into drone technology, with several pioneering initiatives underscoring the growing commitment to this innovative delivery method. Notably, Amazon unveiled its Amazon Prime Air service in late 2013, marking a significant milestone in the commercial application of drones for delivery~\cite{shavarani2018application}. Alibaba has been experimenting with drone deliveries since 2015, demonstrating the feasibility and value of integrating drones into existing logistics frameworks~\cite{han2024value}. In response to the COVID-19 pandemic, UPS initiated a drone delivery service in a Florida retirement community in 2020, showcasing the technology's potential in addressing urgent healthcare needs~\cite{premack2020}. Additionally, DHL made headlines in 2019 by launching its first regular, fully-automated, and intelligent urban drone delivery service, setting a new standard for urban logistics solutions~\cite{dhl2019}. 

\begin{figure}[t]
	\centering
	\includegraphics[width=.97\columnwidth]{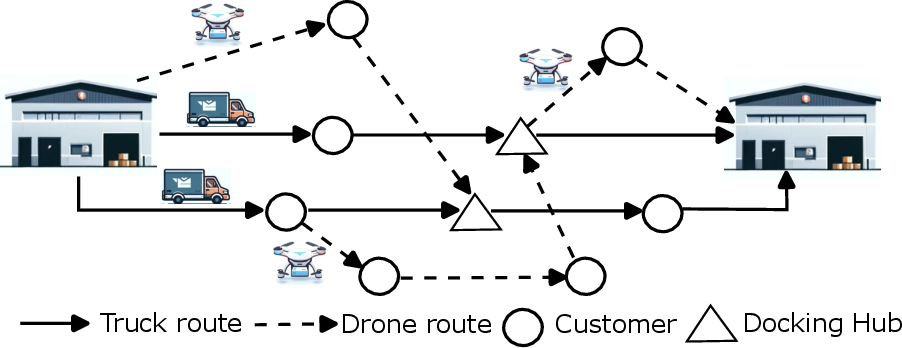}
	\caption {An overview of a truck-drone operation in the vehicle routing problem with drones (VRPD). Trucks and drones collaborate to distribute parcels to customers, with the docking hubs serving as the retrieval points for trucks to collect returning drones.}
	\label{fig:overview}
\end{figure}

The Vehicle Routing Problem with Drones (VRPD) represents an evolution of the traditional Vehicle Routing Problem (VRP), incorporating both trucks and drones into its operational framework. This novel problem was initially introduced by Wang and Sheu~\cite{wang2019vehicle}, distinguishing itself from VRP by utilizing two distinct modes of delivery: trucks, which not only serve customers but also launch drones, and the drones, which are deployed for deliveries from trucks (drones can also be deployed from depots), and subsequently retrieved at designated docking hubs~\cite{wang2019vehicle}. Drones offer the flexibility of being launched and retrieved multiple times at various times and locations. The essence of VRPD lies in optimizing both the delivery routes for trucks and drones as well as the strategic selection of launch and retrieval points for drones, with the goal of minimizing the total operational costs.

Numerous solutions have been developed to tackle VRPD and its variants, each considering a variety of critical factors, including the speed of drones~\cite{tamke2023vehicle}, the number of drones~\cite{zhou2023exact}, drone energy consumption~\cite{xia2023branch}, and the impact on carbon emissions~\cite{kuo2023applying}, among others. However, a notable limitation persists: most of these solutions resort to heuristic or meta-heuristic approaches due to the computational complexity stemming from the NP-Hard nature of the problem. Consequently, these solutions are often constrained to handling only a small number of instances, \emph{i.e.,} the number of customer locations. Moreover, the stochastic nature of heuristic methods, which rely on random solution generation and modification, poses a significant challenge in ensuring high solution quality.

To address the limitations inherent in current heuristic methodologies for VRPD, our approach begins with a thorough assessment of these heuristics found in the literature. We systematically deconstruct these algorithms into universal components. Building upon the dissection of heuristic algorithms, we present SmartPathfinder that provides standardized guidelines for integrating a Reinforcement Learning (RL) framework with the fundamental components of heuristic algorithms aiming to enhance both computational efficiency and solution quality. Specifically, we intertwine the solution space of heuristic algorithms with the decision-making process of an RL agent. This agent explores the solution space under the guidance of a novel reward function designed to optimize both solution quality and computation speed simultaneously, facilitating the computation of solutions for large-scale problems with extensive customer locations. Additionally, our approach incorporates a mechanism for `solution escape', empowering the system to circumvent potential local optima, thereby refining the resilience and efficacy of the novel RL-heuristic algorithm integration. The outcomes of our integration, pairing the RL framework with a state-of-the-art heuristic algorithm for VRPD, demonstrate notable improvements in solution quality and computation speed by up to 28.4\% and 27.3\%, respectively, in comparison with the heuristic algorithm without RL integration. In summary, the contributions of our work can be summarized as follows.

\begin{itemize}
	\item We perform an in-depth examination of heuristic solutions for the Vehicle Routing Problem with Drones (VRPD), breaking them down into their fundamental components for seamless integration with the reinforcement learning (RL) framework.
	\item We introduce novel guidelines for the integration of a RL framework within VRPD heuristic solutions.
	\item We implemented the RL integration for a state-of-the-art VRPD solution, showcasing our methodology's applicability and flexibility.
	\item Through a comprehensive computational study, we highlight the significant advantages of our RL-enhanced solution for VRPD, particularly in terms of computational efficiency and the superior quality of the solutions generated.
\end{itemize}

This paper is organized as follows. In Section~\ref{sec:related_work}, we present a literature review on various solutions for VRPD and its variants. In Section~\ref{sec:preliminaries}, we introduce the precise definition of VRPD and conduct a thorough analysis of existing heuristic approaches for VRPD, setting the stage for explaining how the RL framework is integrated with a heuristic solution. In Section~\ref{sec:proposed_approach}, we present the detailed RL framework design and how it can be incorporated with a heuristic VRPD solution. We then present the evaluation results in Section~\ref{sec:simulation} and conclude in Section~\ref{sec:conclusion}.

\section{Related Work}
\label{sec:related_work}

The goal of VRPD is to optimize the delivery routes for a fleet of trucks and drones, ensuring they operate in a synergistic manner to efficiently distribute parcels to customers~\cite{schermer2019matheuristic}. Specifically, this involves the strategic deployment of drones from either the trucks or depots, enabling them to directly deliver parcels to customers. Upon the completion of their delivery, these drones are designed to reunite with the trucks at predetermined rendezvous points. This section delves into a comprehensive review of state-of-the-art approaches developed for tackling VRPD and its diverse variations, particularly focusing on papers published within five years.

Zhou \emph{et al.} explored a unique variation of VRPD, focusing on the optimization of the number of drones assigned to each vehicle, a contrast to the conventional approach of static drone allocation~\cite{zhou2023exact}. To address this problem, they employed a branch-and-price algorithm with a tabu search strategy. Imran \emph{et al.} marked a pioneering step towards the realization of a drone-based parcel delivery ecosystem, seamlessly incorporating autonomous vehicles (AV) into the VRPD framework~\cite{imran2023vrpd}. Their approach seeks to minimize operational expenses by adeptly selecting AVs from a pool of available AVs through a crowd-sourced methodology. This strategy enables the dynamic assignment of AVs to specific customers, further refining the efficiency of route planning by leveraging real-time traffic data. Both Mixed Integer Linear Programming (MILP) and a greedy algorithm were designed to solve their problem.

Tamke \emph{et al.} considered a variation of VRPD, dubbed the Vehicle Routing Problem with Drones and Drone Speed Selection (VRPD-DSS)~\cite{tamke2023vehicle}. Their model prioritizes the minimization of operational expenses while accommodating the selection of discrete speed settings for drones. Their findings reveal that adjusting drone speeds according to specific delivery requirements offers significant cost reductions over traditional methodologies that do not account for speed variability. Xia \emph{et al.} enhanced the VRPD framework by introducing drone stations that are mainly used for collection, storage and recharge of drones as well as replenishing parcels for both trucks and drones~\cite{xia2023branch} . The drone stations serve to bolster the synergy between trucks and drones while addressing the impact of varying load weights on drone energy consumption~\cite{xia2023branch}. They designed an advanced branch-and-price-and-cut algorithm taking into account the drone stations. 

The authors of~\cite{kuo2023applying} delve into a variant of VRPD with dual objectives of minimizing delivery times and reducing carbon emissions. Leveraging the Non-Dominated Sorting Genetic Algorithm II (NSGA-II), their approach adeptly balances these objectives. Their findings demonstrated the significant potential of drones to enhance both environmental sustainability and delivery efficiency. Meanwhile, the research presented in~\cite{yin2023robust} targets the application of VRPD within the context of disaster relief, acknowledging the inherent uncertainties in demand and travel times that characterize such scenarios. By formulating an enhanced branch-and-price-and-cut (BPC) algorithm, this study uncovers efficient routing strategies that underscore the substantial advantages of integrating trucks with drones over traditional truck-only methods, suggesting its potential for real-world disaster response.

\begin{table*}[!t] 
	\centering
	\caption{Decomposition of heuristic algorithms for VRPD into four universal components.}
	\begin{tabularx}{\textwidth}{ | l | X | c | c | c | c | }
		\hline
		Papers & Heuristic & Initialization & Solution Modification & Shuffling & Solution Evaluation \\[6pt] 
		\hline
		\cite{kuo2022vehicle} & Variable Neighborhood Search (VNS) & \checkmark & \checkmark  & \checkmark  & \checkmark \\[6pt]  
		\hline
		\cite{sacramento2019adaptive}~\cite{momeni2023new} & Largest/Nearest Neighborhood Search (LNS/NNS) & \checkmark & \checkmark &   & \checkmark \\[6pt] 
		\hline
		\cite{mara2023solving}~\cite{kuo2023applying} & Genetic/Memetic Algorithm (GA/MA) & \checkmark & \checkmark  & \checkmark  & \checkmark \\[6pt]
		\hline
		\cite{huang2022solving} & Ant Colony Optimization (ACO) & \checkmark & \checkmark &  & \checkmark \\[6pt] 
		\hline
		\cite{han2020metaheuristic}~\cite{lei2022dynamical} & Artificial Bee Colony (ABC) & \checkmark & \checkmark & \checkmark & \checkmark \\[6pt] 
		\hline
		\cite{liu2020two}~\cite{meng2023multi} & Simulated Annealing (SA) & \checkmark & \checkmark &  & \checkmark \\[6pt] 
		\hline
		\cite{imran2023vrpd} & Greedy Algorithm & \checkmark &  &  & \checkmark \\[6pt]
		\hline
		\cite{xia2023branch}~\cite{yin2023robust} & Branch-and-Price-and-Cut algorithm (BPC) & \checkmark & \checkmark &  & \checkmark \\[6pt]
		\hline
	\end{tabularx}
	\label{table:soa_heuristics}
\end{table*}

Mara \emph{et al.}~\cite{mara2023solving} introduced the Electric Vehicle-Drone Routing Problem (EVDRP), where drones collaborate with electric vehicles for last-mile deliveries. They minimize the total completion time, considering EV battery limitations and recharging stations using the memetic algorithm with problem-specific operators to efficiently solve the problem. The authors in~\cite{meng2023multi} proposed a new model for solving VRPD, allowing drones to perform multiple pickups and deliveries per flight. They develop  a two-stage heuristic and show that allowing multiple visits and combined pickup/delivery significantly reduces the total cost compared to traditional approaches. Momeni \emph{et al.}~\cite{momeni2023new} proposed a novel VRPD model that minimizes delivery time while considering drone energy consumption at varying altitudes. They developed a two-phase heuristic algorithm based on nearest neighborhood and local search and demonstrated the model's effectiveness in optimizing routes, particularly for deliveries at different heights, for both small, large and real-world instances.

Previous studies have developed effective solutions for numerous variants of VRPD, with a significant emphasis on heuristic strategies owing to the NP-Hard complexity of the problem. However, the complexity of the problem limits such approaches, typically confining them to a relatively small scope of instances in terms of the number of customer locations. Additionally, the inherently probabilistic characteristic of heuristic algorithms, dependent on randomly generating and adjusting solutions, represents a notable obstacle in achieving consistently superior solution quality. In contrast, SmartPathfinder uniquely employs a novel machine learning-driven approach through seamless integration with heuristic algorithms to address the challenges. 

\section{Preliminaries}
\label{sec:preliminaries}

\subsection{Definition of VRPD}
\label{sec:definition_of_vrpd}

The Vehicle Routing Problem with Drones (VRPD) is defined in a graph $G = (N,A)$, where $N$ denotes a collection of nodes, including a depot node, customer nodes $C = \{c_1, c_2, ..., c_n\}$, and docking hub nodes $O = \{o_1, o_2, ..., o_m\}$. $A$ denotes the arcs connecting these nodes. Vehicles and drones in this scenario are denoted by a set $K$ and $D$, respectively. Drones are characterized by their constrained payload capacity of up to $L^D$ weight units and the operational range of up to $T^D$. Drones benefit from the ability to navigate more direct paths, facilitating expedited deliveries to customers. Conversely, trucks can accommodate a more substantial cargo, up to $L^R$ drones and $L^T$ weight units of parcels, albeit they are subject to the limitations imposed by the road network. A parcel destined for a customer $i \in C$, weighing $g_i$, must not exceed $L^T$, \emph{i.e.,} it qualifies for drone delivery if $g_i \leq L^D$. The travel times for trucks and drones from node $i \in N$ to $j \in N$ are denoted by $t_{ij}^T$ and $t_{ij}^D$ respectively, with drone travel time $t_{ij}^D$ invariably being less than that of trucks $t_{ij}^T$, showcasing their advantage in speed and efficiency.

Drones may serve customers independently within their flight range, or be transported by trucks to customers located outside this range. It is assumed that the battery-swapping and drone-loading times are negligible. Additionally, the following variables are defined to set up the objective function.

\begin{itemize}
	\item $x_{ijk}$: Equals $1$ if the $k$-th truck travels arc $(i, j) \in A$ independently, and $0$ otherwise.
	\item $y_{ijd}$: Equals $1$ if the $d$-th drone travels arc $(i, j) \in A$ independently, and $0$ otherwise.
	\item $u_{ijk}$: Equals $1$ as long as the $k$-th truck carries one or more drones through arc $(i, j) \in A$, and $0$ otherwise.
\end{itemize}

The objective of VRPD is to minimize the total operation cost, comprised of the fixed truck employment cost $F^T$ and the variable transportation costs for both trucks $C^T$ and drones $C^D$, per unit travel time which is defined as follows.

\begin{align*}
	\min \quad \Biggl[ F^T \biggl( \sum_{(i,j) \in A} \sum_{k \in K} x_{ijk} + \sum_{(i,j) \in A} \sum_{k \in K} u_{ijk} \biggl) +\\ C^T\sum_{(i,j) \in A} \sum_{k \in K} t_{ij}^T(x_{ijk} + u_{ijk}) + C^D\sum_{(i,j) \in A} \sum_{d \in D} t_{ij}^D y_{ijd} \Biggl].    \label{eq:vrpd_obj}
\end{align*}

The objective function, as outlined in~\cite{wang2019vehicle}, is subject to various constraints to ensure feasibility and efficiency in operations involving trucks and drones. Key among these constraints are two that guarantee both trucks and drones return to their starting depot. Another pair of constraints ensures that each customer node is visited exactly once, meaning there's only one incoming and outgoing connection for each node. Additionally, at every docking station, the number of arriving and departing drones is kept equal to maintain balance. To prevent overloading, two constraints limit the truck from carrying more drones than its capacity allows. Moreover, three specific constraints are in place to regulate the maximum flight duration of drones, ensuring their operations are within feasible limits. There are also five more constraints focusing on the parcel-carrying capacities of both drones and trucks to prevent overloading. Detailed mathematical formulations of these constraints are available in~\cite{wang2019vehicle}.

\subsection{Analysis of Heuristic Algorithms for VRPD}
\label{sec:dissection}

In this section, we perform an analysis of heuristic methodologies applied to solving VRPD, aiming to decompose these algorithms into their fundamental elements for incorporation with the RL framework. This analysis encompasses a comprehensive review of 13 papers  that introduce heuristic solutions for VRPD, published between 2019 and 2023. As illustrated in Table I, despite the diverse operational mechanisms of these algorithms, our extensive analysis has identified four core components common to many heuristic approaches: Solution Initialization, Solution Modification, Solution Evaluation, and Solution Shuffling.

The Solution Initialization component is pivotal in establishing the solution structure and creating the baseline solutions upon which heuristic algorithms iteratively improve. The Solution Modification component stands as the crux of the algorithmic process, employing a tailored set of rules to iteratively refine the solution. In the Solution Evaluation component, the focus is on assessing the solution's efficacy, specifically measuring its impact on minimizing the total operational expenses incurred in servicing all customer locations. It is noteworthy that while the aforementioned components are universally present across most heuristics, the Solution Shuffling mechanism—which involves the random rearrangement of solution elements to avert stagnation at local optima—is not as uniformly integrated. This systematic breakdown of heuristic methodologies facilitates the seamless integration of the RL framework with the components of the heuristics. A detailed explanation of this integration process is explained in the subsequent section.

\section{Design of SmartPathfinder}
\label{sec:proposed_approach}

\subsection{Overview}
\label{sec:overview}

\begin{figure}[t]
	\centering
	\includegraphics[width=.98\columnwidth]{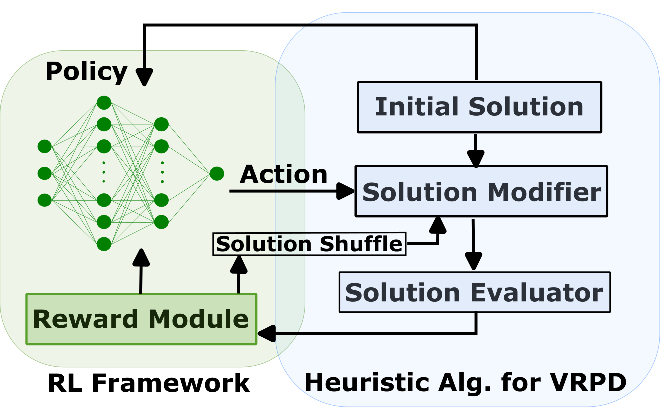}
	\caption {An architecture of SmartPathfinder, illustrating integration with the RL framework with a heuristic algorithm for VRPD.}
	\label{fig:overview}
\end{figure}

Building on the universal components of heuristic algorithms for VRPD, SmartPathfinder aims to seamlessly incorporate a RL framework within the structure of a heuristic algorithm to improve both the quality of solutions and computational efficiency. Fig.~\ref{fig:overview} depicts the architecture of SmartPathfinder. As shown, it begins with an initial solution, which activates the RL framework's policy network. The policy network, once initialized, outputs actions that are forwarded to the solution modifier. These actions, designed in accordance with the underlying heuristic algorithm, dictate the modification of the initial solution. Following this, the solution evaluator assesses the modified solution's efficacy. Subsequently, the evaluation scores of both the modified and the original solutions are relayed to the RL framework's reward module, where a reward is computed based on these evaluations. In particular, if the solution fails to improve after a predetermined number of iterations, the procedure diverts from policy network updating to solution shuffling. This involves rearranging the solution before subjecting it to reevaluation to avoid potential local optima. In situations where an improvement in the solution is observed, the reward module feeds the results of the evaluation back into the policy network for updates. The updated network then issues new actions aimed at further refining the solution. This cycle of solution evaluation, reward assessment, and solution modification iterates, fostering continuous improvement in the solution quality.

\subsection{Action Space}
\label{sec:action_space}

The design of the action space is tailored to the solution modification capabilities inherent to heuristic algorithms, \emph{i.e.,} each action represents a solution alteration method for a specific heuristic algorithm. Unlike the probabilistic solution alteration typical in heuristic algorithms~\cite{kuo2022vehicle}\cite{mara2023solving}\cite{huang2022solving}\cite{meng2023multi}, this approach advocates for a strategic selection of actions driven by reward feedback, aiming for solutions of superior quality. For instance, key operations from a genetic algorithm (GA)-based approach to solving VRPD~\cite{mara2023solving} such as the parent selection, crossover, and mutation, are integrated as specific actions within the RL framework. This integration allows for more strategic action selection, guided by the RL's policy network. Similarly, the core operations of the Neighborhood Search (NS) algorithm for VRPD~\cite{kuo2022vehicle}, which include neighborhood moves essential for local search, such as node swaps, entire swaps, node insertion, whole insertion, node reversal, entire reversal, sortie removal, and sortie addition, are also adapted as actionable strategies within the RL framework. 

\subsection{State Space}
\label{sec:state_space}

The state space is a critical construct that defines all necessary information for an agent to make informed decisions regarding its actions. More specifically, the state space of SmartPathfinder includes information related to both the quality of potential solutions and the efficiency of computational processes. Solution quality is gauged through the solution evaluation outcomes at any given time step $t$, denoted by $\mathcal{S}_t$. This metric is derived from the heuristic algorithm's solution evaluator, providing a snapshot of the current solution's effectiveness. Additionally, the evaluation result of the solution prior to the current one is denoted by $\mathcal{S}_{t-1}$, offering a comparative perspective on solution progression or regression. On the other hand, computational efficiency is quantified by the cumulative actions executed by the agent up to time step $t$, denoted by $\mathcal{A}_t$. This measurement reflects the agent's operational speed and efficiency, which are especially important in environments where computational resources or time are constrained. To navigate the trade-offs between solution quality and computational speed, the state space also integrates performance weights $w_1$ and $w_2$. These weights are utilized to balance and prioritize these performance dimensions, guiding the agent towards a more effective decision-making process. 

\subsection{Reward Function}
\label{sec:reward_function}

The reward function of SmartPathfinder is designed to simultaneously enhance the solution quality and minimize computational time, embodying a dual objective within the RL framework. Furthermore, it offers configurability, allowing users to tailor the weighting between solution quality and computational efficiency according to specific needs. To quantitatively evaluate solution quality, the reward function employs the net improvement of the solution, calculated as $\mathcal{S}_{t} - \mathcal{S}_{t-1}$. This formulation inherently associates a negative reward with any deterioration in solution quality, incentivizing progress and penalizing regression. Also, the function prioritizes computational speed by incorporating the total number of actions executed by the agent, \emph{i.e.,} by assigning a negative reward as the count of actions increases. Consequently, with these considerations, the reward function $R_t$ at time step $t$ is defined as follows. 

\begin{displaymath}
	R_t = w_1(\mathcal{S}_{t} - \mathcal{S}_{t-1}) - w_2\mathcal{A}_t,
\end{displaymath}

\noindent where $w_1$ is the weighting parameter for the solution quality, $w_2$ is the weighting parameter for the computational time, and $\mathcal{A}_t$ is the total number actions taken.

\section{Evaluation Results}
\label{sec:simulation}

\subsection{Evaluation Setup}

\begin{figure}[h!]
	\centering
	\includegraphics[width=.95\columnwidth]{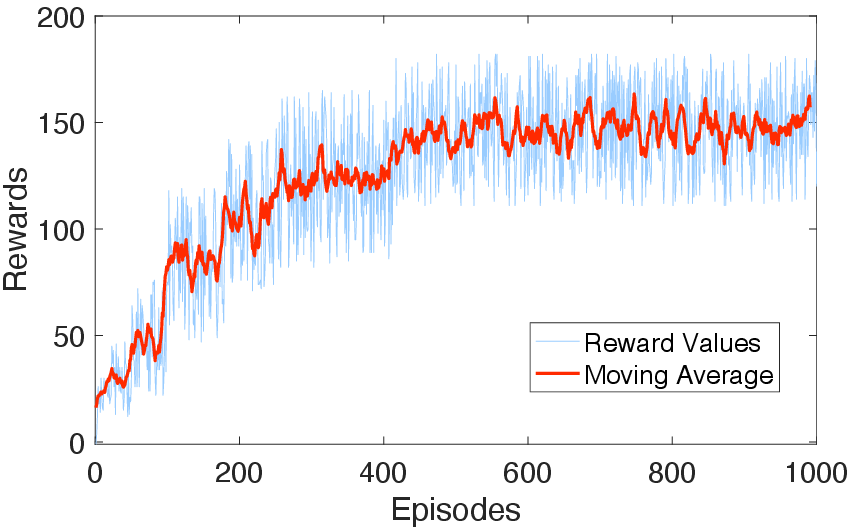}
	\caption {The reward values of the memetic algorithm-based heuristic solution integrated with the RL framework, illustrating the convergence of the reward value.}
	\label{fig:convergence}
\end{figure}

To evaluate the performance of SmartPathfinder, we implemented the integration of the RL framework with a state-of-the-art heuristic algorithm designed based on the memetic algorithm~\cite{mara2023solving}. Fig.~\ref{fig:convergence} demonstrates the convergence of the reward value for SmartPathfinder integrated with the heuristic algorithm~\cite{mara2023solving}. Additionally, for a more effective performance comparison, we implemented another heuristic algorithm based on neighborhood search~\cite{kuo2022vehicle}. Throughout this section, the memetic algorithm-based solution is referred to as MA, the neighborhood search-based approach as NS, and the RL-enhanced method as RL+MA. All implementation was performed on a computer equipped with an AMD Ryzen 7 7840HS CPU, 16GB RAM, NVIDIA GeForce RTX 4050 GPU, and Windows 11. The code was written in Python 3.10.

Our evaluation hinges on two primary metrics: solution quality and computational efficiency. The quality of the solution is measured by the total operational time, which is the total amount of time needed to serve all customer locations. This measure has been widely adopted by numerous heuristic algorithms for VRPD for performance evaluation~\cite{tamke2021branch, kuo2023applying, schermer2019matheuristic}. Computational efficiency, on the other hand, is gauged by the elapsed time from the initiation of the algorithm to the moment the final solution is derived. We evaluate the performance of our proposed solution across various scenarios, characterized by differing numbers of customer nodes by up to 100 customer nodes. For distance calculations, we adopt the Manhattan distance metric for truck movements and the Euclidean distance for drone operations, similar to the methods used in many VRPD heuristic solutions~\cite{mara2023solving, meng2023multi, sacramento2019adaptive}. We adopted a dataset from ~\cite{kuo2022vehicle} for performance evaluation. The data set comprises of 72 test instances with varying numbers of customer nodes. For fair performance comparison, we excluded the customer time windows from the dataset.

\begin{figure*}[h]
	\centering
	\includegraphics[width=.99\textwidth]{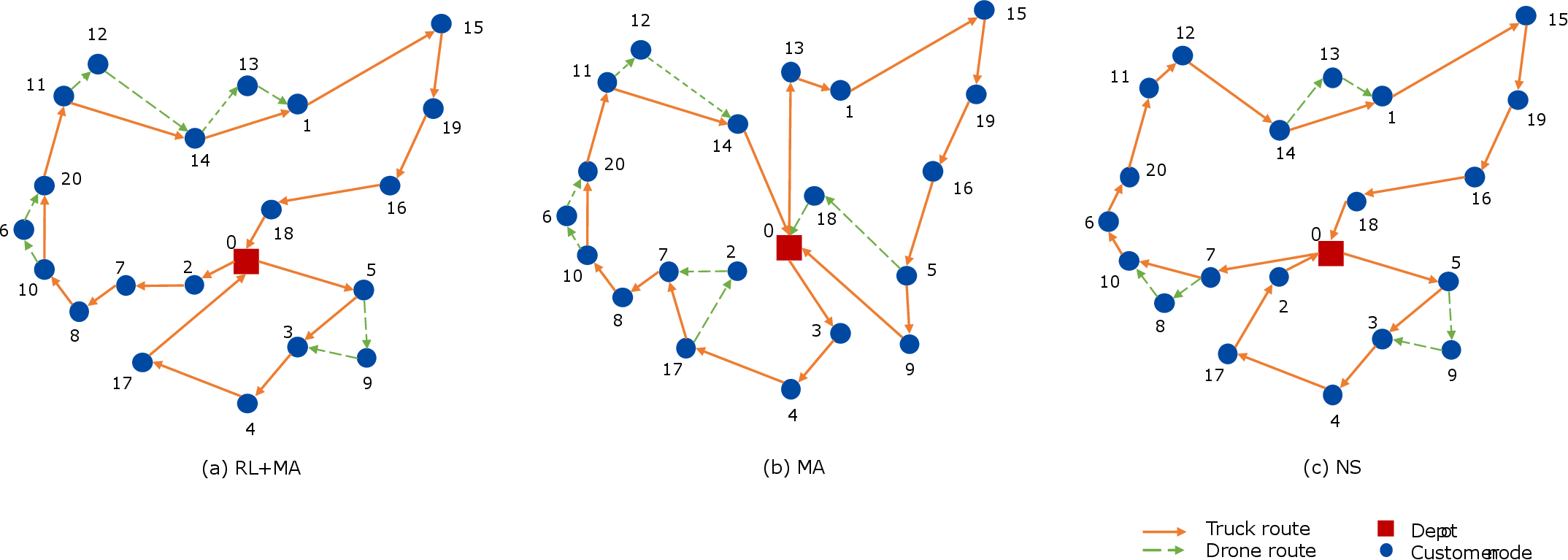}
	\caption {An example solution generated with (a) RL+MA, (b) MA, and (c) NS. The integration of the RL framework with MA results in more efficient paths for both trucks and drones compared with MA and NS.}
	\label{fig:route_example}
\end{figure*}

In configuring the memetic algorithm, we adhere to the parameter settings detailed by Mara \emph{et al.}~\cite{mara2023solving}. In particular, our evaluation method does not assume electric vehicles (EVs); hence, we employ the same neighborhood move strategies as those outlined in Kuo \emph{et al.}~\cite{kuo2022vehicle} and Mara \emph{et al.}~\cite{mara2023solving}, with the exception of EV-specific maneuvers such as recharging insertion, change station, moving station, and remove station.

\subsection{Solution Quality}

\begin{figure}[h]
	\centering
	\includegraphics[width=.9\columnwidth]{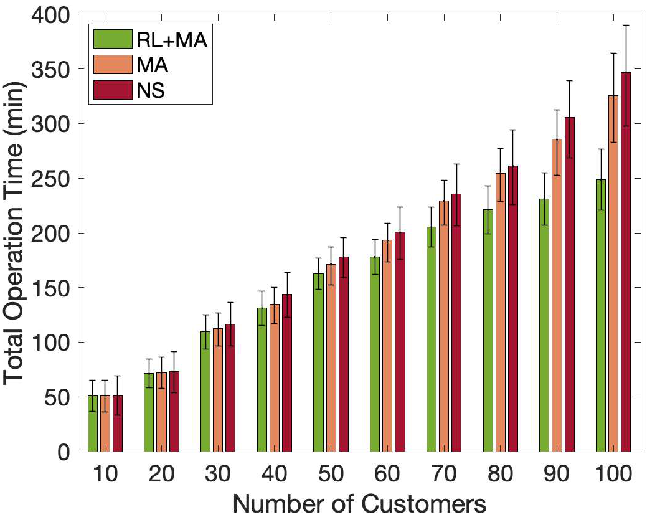}
	\caption {The solution quality with varying numbers of customers. RL+MA significantly improves the solution quality especially for large-scale problems.}
	\label{fig:num_neighbors}
\end{figure}

We compare the total operational time across three different strategies: RL+MA, MA, and NS. The total operational time was measured by varying the customer node counts, with all customer nodes being randomly deployed within the target area. The findings, illustrated in Fig.~\ref{fig:num_neighbors}, demonstrate that RL+MA outperforms MA and NS, particularly as the complexity of the problem increases with more customer nodes. More specifically, initially, when the number of customer nodes is low, the distinction in performance between the RL-enhanced strategy and conventional methods is minimal. This trend is attributed to the constrained solution space, which allows all evaluated algorithms to relatively easily identify high-quality solutions. Nonetheless, the advantage of integrating RL becomes increasingly apparent with the growth in customer numbers. Specifically, for scenarios involving 100 customer nodes, the RL-enhanced strategy reduces the total operational time by up to 23.7\% compared to MA, and by 28.4\% relative to NS. These results highlight the significant benefits of incorporating RL into heuristic algorithms for tackling VRPD, especially in more demanding scenarios with a larger customer base.

Fig.~\ref{fig:route_example} presents example solutions created using RL+MA, MA, and NS algorithms. Interestingly, even with a modest customer base of 20, each method yields distinct solutions with varying levels of efficiency in terms of total operational time. Notably, the NS algorithm produces a solution with only 3 drone sorties, in contrast to the 4 drone sorties observed in the RL+MA and MA solutions. This more efficient drone usage contributes to a reduction in overall operational time. Additionally, it is observed that although RL+MA and MA both deploy the same two trucks, the routes taken by these trucks differ significantly. Such variations likely stem from the RL's ability to iteratively refine routing decisions through its adaptive learning mechanism, optimizing routes based on trial and error.

\subsection{Computation Time}

\begin{figure}[h]
	\centering
	\includegraphics[width=.9\columnwidth]{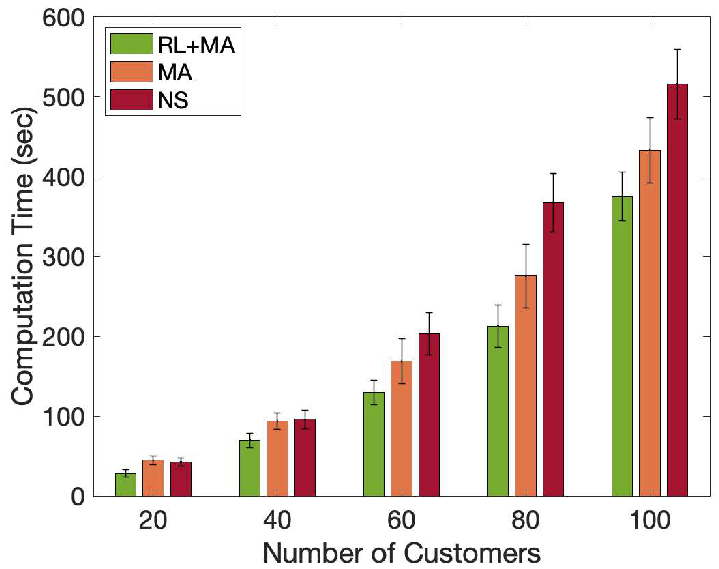}
	\caption {The computation time with varying numbers of customers. RL+MA significantly improves the computation time across all test instances.}
	\label{fig:execution_time}
\end{figure}

Another crucial part of our analysis focuses on assessing how incorporating the RL framework influences computational efficiency, especially when compared to MA and NS. The results are depicted in Fig.~\ref{fig:execution_time}, demonstrating that the RL-enhanced approach, RL+MA, consistently achieves reductions in computation time across various problem sizes, even in scenarios with a relatively small number of customers. A key observation from our findings is that the RL framework not only maintains its efficiency advantage across all tested scales but also that this advantage becomes more significant as the number of customers increases. For instance, in cases involving 100 customers, the integration of RL leads to a decrease in computation time by approximately 13.2\% compared to MA, and an even more substantial 27.3\% compared to NS. This performance difference highlights the advantage of leveraging machine learning techniques to explore the solution space more effectively, as opposed to traditional methods that rely more on heuristic or stochastic solution adjustments.

A notable advantage of the RL+MA approach is its capability to tackle considerably larger problems, which were previously beyond the reach of existing heuristic methods, not only because of the substantial decrease in computation time, but also the more effective exploration of the solution space. Specifically, with RL+MA, we were able to solve problems with up to 200 customers in just an average of 1,121 seconds. The result is particularly remarkable, considering that the existing benchmarks in the literature limit evaluations to a maximum of 100 customers for MA and 50 customers for NS.

\subsection{Ablation Study}
\label{sec:ablation_study}

A key feature of SmartPathfinder is its ability to skip actions and randomly shuffle the solution if it doesn't show any improvement after a set number of $k$ attempts. This technique is specifically engineered to avoid falling in local optima, thereby optimizing the solution discovery process. To systematically evaluate the influence of parameter $k$ on the algorithm's efficiency and effectiveness, we undertook an ablation study. The objective of this study was to discern the impact of varying $k$ values on the solution quality in terms of the operational time and the computational delay. 

\begin{figure}[h]
	\centering
	\begin{minipage}{.49\columnwidth}
		\centering
		\includegraphics[width=\textwidth]{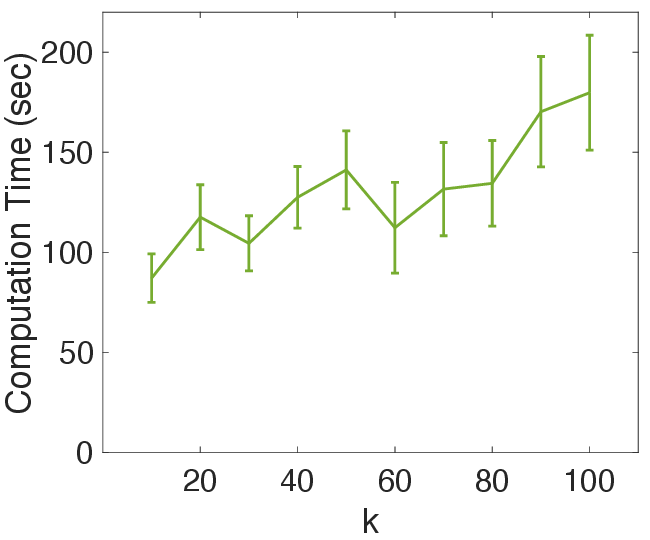}
		\vspace{-15pt}
		\caption {The computation time measured with varying threshold $k$.}
		\label{fig:time_per_k}
	\end{minipage}%
	\hspace*{1mm}
	\begin{minipage}{.49\columnwidth}
		\centering
		\includegraphics[width=\textwidth]{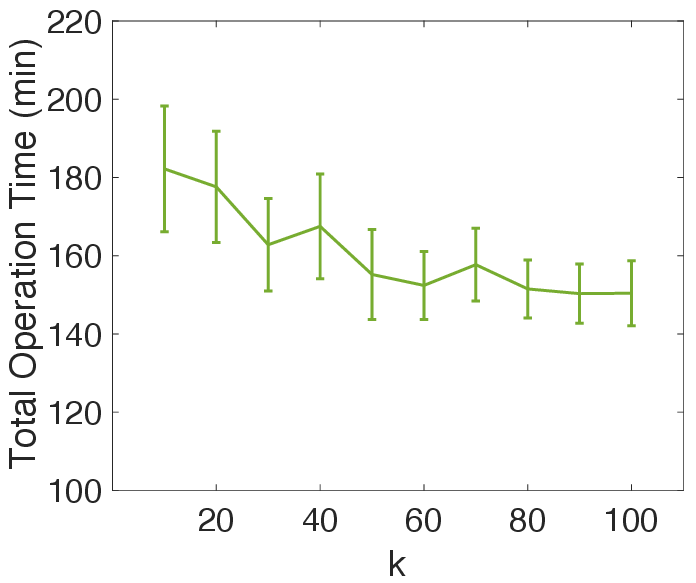}
		\vspace{-15pt}
		\caption {The solution quality measured with varying threshold $k$.}
		\label{fig:performance_per_k}
	\end{minipage}
\end{figure}


First, We adjusted the value of $k$ and tracked how it influenced the computation time. The results are detailed in Fig.~\ref{fig:time_per_k}, providing guidance for selecting an optimal $k$ value. The results show that as the value of $k$ increases, a notable increase in computation time is observed. This phenomenon can be attributed to the fact that a higher $k$ value permits a more extensive examination of the solution space, as it delays the algorithm's exit from local optima in pursuit of potentially superior solutions. Consequently, this expanded exploration results in an increased computation time. Upon initial examination, the results appear to indicate that choosing a smaller $k$ value could be beneficial for enhancing computational efficiency.


However, as depicted in Fig.~\ref{fig:performance_per_k}, an improvement in performance, specifically regarding total operational time, is noted with an increase in $k$. This phenomenon highlights a discernible trade-off between computation time and overall performance when determining the optimal $k$ value. Notably, the curve representing total operational time exhibits a plateau beyond a certain $k$ value, which, in the context of this simulation, occurs approximately at $k=60$. This observation implies that choosing a $k$ value at the onset of this plateau phase—where additional increases in $k$ no longer significantly enhance performance—presents a strategic approach for optimizing operational efficiency.

\section{Conclusion}
\label{sec:conclusion}

We have presented SmartPathfinder, a novel approach to seamlessly integrate a RL framework with heuristic solutions for VRPD, targeting enhancements in both the solution quality and computational efficiency. This novel integration is facilitated by a thorough analysis and decomposition of heuristic solutions into universal components, followed by the design of the RL framework, and
reassembly of the components seamlessly incorporating the RL framework. To evaluate the effectiveness of SmartPathfinder, we implemented a state-of-the-art heuristic solution for VRPD integrated with the RL framework and demonstrated significant improvements in both the effectiveness of the solutions and reduction in computational time. We believe that this pioneering effort marks a significant stride toward refining and accelerating the decision-making process in drone-assisted delivery systems. 

As our future work, we aim to explore the theoretical limits of solution quality improvement achievable through integrating heuristic solutions with our RL framework. Additionally, we plan to investigate the adaptability of our proposed RL integration approach across various optimization problems, assessing its effectiveness with heuristic solutions tailored for different challenges. Lastly, we will conduct a broader assessment of SmartPathfinder by testing it with a diverse array of heuristic algorithms.

\bibliographystyle{IEEEtran}
\bibliography{mybibfile}

\begin{thebibliography}{10}
\providecommand{\url}[1]{#1}
\csname url@samestyle\endcsname
\providecommand{\newblock}{\relax}
\providecommand{\bibinfo}[2]{#2}
\providecommand{\BIBentrySTDinterwordspacing}{\spaceskip=0pt\relax}
\providecommand{\BIBentryALTinterwordstretchfactor}{4}
\providecommand{\BIBentryALTinterwordspacing}{\spaceskip=\fontdimen2\font plus
\BIBentryALTinterwordstretchfactor\fontdimen3\font minus
  \fontdimen4\font\relax}
\providecommand{\BIBforeignlanguage}[2]{{%
\expandafter\ifx\csname l@#1\endcsname\relax
\typeout{** WARNING: IEEEtran.bst: No hyphenation pattern has been}%
\typeout{** loaded for the language `#1'. Using the pattern for}%
\typeout{** the default language instead.}%
\else
\language=\csname l@#1\endcsname
\fi
#2}}
\providecommand{\BIBdecl}{\relax}
\BIBdecl

\bibitem{agatz2018optimization}
N.~Agatz, P.~Bouman, and M.~Schmidt, ``Optimization approaches for the
  traveling salesman problem with drone,'' \emph{Transportation Science},
  vol.~52, no.~4, pp. 965--981, 2018.

\bibitem{lee2022congestion}
S.~Lee, D.~Hong, J.~Kim, D.~Baek, and N.~Chang, ``Congestion-aware multi-drone
  delivery routing framework,'' \emph{IEEE Transactions on Vehicular
  Technology}, vol.~71, no.~9, pp. 9384--9396, 2022.

\bibitem{goodchild2018delivery}
A.~Goodchild and J.~Toy, ``Delivery by drone: An evaluation of unmanned aerial
  vehicle technology in reducing co2 emissions in the delivery service
  industry,'' \emph{Transportation Research Part D: Transport and Environment},
  vol.~61, pp. 58--67, 2018.

\bibitem{chiang2019impact}
W.-C. Chiang, Y.~Li, J.~Shang, and T.~L. Urban, ``Impact of drone delivery on
  sustainability and cost: Realizing the uav potential through vehicle routing
  optimization,'' \emph{Applied energy}, vol. 242, pp. 1164--1175, 2019.

\bibitem{dorling2016vehicle}
K.~Dorling, J.~Heinrichs, G.~G. Messier, and S.~Magierowski, ``Vehicle routing
  problems for drone delivery,'' \emph{IEEE Transactions on Systems, Man, and
  Cybernetics: Systems}, vol.~47, no.~1, pp. 70--85, 2016.

\bibitem{pugliese2020using}
L.~D.~P. Pugliese, F.~Guerriero, and G.~Macrina, ``Using drones for parcels
  delivery process,'' \emph{Procedia Manufacturing}, vol.~42, pp. 488--497,
  2020.

\bibitem{shavarani2018application}
S.~M. Shavarani, M.~G. Nejad, F.~Rismanchian, and G.~Izbirak, ``Application of
  hierarchical facility location problem for optimization of a drone delivery
  system: a case study of amazon prime air in the city of san francisco,''
  \emph{The International Journal of Advanced Manufacturing Technology},
  vol.~95, pp. 3141--3153, 2018.

\bibitem{han2024value}
B.~R. Han, M.~Li, Y.~Zhang, and P.~Li, ``Value of autonomous last-mile
  delivery: Evidence from alibaba,'' \emph{Available at SSRN}, 2024.

\bibitem{premack2020}
{Premack, Rachel}, ``{America’s largest retirement community can soon receive
  their prescriptions from CVS via a UPS Drone Delivery Service},''
  \url{https://www.businessinsider.com/ups-cvs-drone-deliveries-the-villages-florida-2020-4},
  online; accessed 7 March 2024.

\bibitem{dhl2019}
{DHL}, ``{DHL launches its first regular fully-automated and intelligent urban
  drone delivery service},''
  \url{https://group.dhl.com/en/media-relations/press-releases/2019/dhl-launches-its-first-regular-fully-automated-and-intelligent-urban-drone-delivery-service.html},
  online; accessed 7 March 2024.

\bibitem{wang2019vehicle}
Z.~Wang and J.-B. Sheu, ``Vehicle routing problem with drones,''
  \emph{Transportation research part B: methodological}, vol. 122, pp.
  350--364, 2019.

\bibitem{tamke2023vehicle}
F.~Tamke and U.~Buscher, ``The vehicle routing problem with drones and drone
  speed selection,'' \emph{Computers \& Operations Research}, vol. 152, p.
  106112, 2023.

\bibitem{zhou2023exact}
H.~Zhou, H.~Qin, C.~Cheng, and L.-M. Rousseau, ``An exact algorithm for the
  two-echelon vehicle routing problem with drones,'' \emph{Transportation
  Research Part B: Methodological}, vol. 168, pp. 124--150, 2023.

\bibitem{xia2023branch}
Y.~Xia, W.~Zeng, C.~Zhang, and H.~Yang, ``A branch-and-price-and-cut algorithm
  for the vehicle routing problem with load-dependent drones,''
  \emph{Transportation Research Part B: Methodological}, vol. 171, pp. 80--110,
  2023.

\bibitem{kuo2023applying}
R.~Kuo, E.~Edbert, F.~E. Zulvia, and S.-H. Lu, ``Applying nsga-ii to vehicle
  routing problem with drones considering makespan and carbon emission,''
  \emph{Expert Systems with Applications}, vol. 221, p. 119777, 2023.

\bibitem{schermer2019matheuristic}
D.~Schermer, M.~Moeini, and O.~Wendt, ``A matheuristic for the vehicle routing
  problem with drones and its variants,'' \emph{Transportation Research Part C:
  Emerging Technologies}, vol. 106, pp. 166--204, 2019.

\bibitem{imran2023vrpd}
N.~M. Imran, S.~Mishra, and M.~Won, ``A-vrpd: Automating drone-based last-mile
  delivery using self-driving cars,'' \emph{IEEE Transactions on Intelligent
  Transportation Systems}, vol.~24, no.~9, pp. 9599--9612, 2023.

\bibitem{yin2023robust}
Y.~Yin, Y.~Yang, Y.~Yu, D.~Wang, and T.~Cheng, ``Robust vehicle routing with
  drones under uncertain demands and truck travel times in humanitarian
  logistics,'' \emph{Transportation Research Part B: Methodological}, vol. 174,
  p. 102781, 2023.

\bibitem{kuo2022vehicle}
R.~Kuo, S.-H. Lu, P.-Y. Lai, and S.~T.~W. Mara, ``Vehicle routing problem with
  drones considering time windows,'' \emph{Expert Systems with Applications},
  vol. 191, p. 116264, 2022.

\bibitem{sacramento2019adaptive}
D.~Sacramento, D.~Pisinger, and S.~Ropke, ``An adaptive large neighborhood
  search metaheuristic for the vehicle routing problem with drones,''
  \emph{Transportation Research Part C: Emerging Technologies}, vol. 102, pp.
  289--315, 2019.

\bibitem{momeni2023new}
M.~Momeni, S.~Mirzapour Al-e Hashem, and A.~Heidari, ``A new truck-drone
  routing problem for parcel delivery by considering energy consumption and
  altitude,'' \emph{Annals of Operations Research}, pp. 1--47, 2023.

\bibitem{mara2023solving}
S.~T.~W. Mara, R.~Sarker, D.~Essam, and S.~Elsayed, ``Solving electric
  vehicle--drone routing problem using memetic algorithm,'' \emph{Swarm and
  Evolutionary Computation}, vol.~79, p. 101295, 2023.

\bibitem{huang2022solving}
S.-H. Huang, Y.-H. Huang, C.~A. Blazquez, and C.-Y. Chen, ``Solving the vehicle
  routing problem with drone for delivery services using an ant colony
  optimization algorithm,'' \emph{Advanced Engineering Informatics}, vol.~51,
  p. 101536, 2022.

\bibitem{han2020metaheuristic}
Y.-q. Han, J.-q. Li, Z.~Liu, C.~Liu, and J.~Tian, ``Metaheuristic algorithm for
  solving the multi-objective vehicle routing problem with time window and
  drones,'' \emph{International Journal of Advanced Robotic Systems}, vol.~17,
  no.~2, p. 1729881420920031, 2020.

\bibitem{lei2022dynamical}
D.~Lei, Z.~Cui, and M.~Li, ``A dynamical artificial bee colony for vehicle
  routing problem with drones,'' \emph{Engineering Applications of Artificial
  Intelligence}, vol. 107, p. 104510, 2022.

\bibitem{liu2020two}
Y.~Liu, Z.~Liu, J.~Shi, G.~Wu, and W.~Pedrycz, ``Two-echelon routing problem
  for parcel delivery by cooperated truck and drone,'' \emph{IEEE Transactions
  on Systems, Man, and Cybernetics: Systems}, vol.~51, no.~12, pp. 7450--7465,
  2020.

\bibitem{meng2023multi}
S.~Meng, X.~Guo, D.~Li, and G.~Liu, ``The multi-visit drone routing problem for
  pickup and delivery services,'' \emph{Transportation Research Part E:
  Logistics and Transportation Review}, vol. 169, p. 102990, 2023.

\bibitem{tamke2021branch}
F.~Tamke and U.~Buscher, ``A branch-and-cut algorithm for the vehicle routing
  problem with drones,'' \emph{Transportation Research Part B: Methodological},
  vol. 144, pp. 174--203, 2021.

\end{thebibliography}

\end{document}